\documentclass[10pt,letterpaper]{article}

\usepackage[margin=0.75in]{geometry}

\usepackage{graphicx}

\usepackage{amsmath,amssymb}

\usepackage{changepage}

\usepackage[utf8x]{inputenc}

\usepackage{textcomp,marvosym}

\usepackage{cite}

\usepackage{nameref,hyperref}

\usepackage{microtype}
\DisableLigatures[f]{encoding = *, family = * }

\usepackage[table]{xcolor}

\usepackage{array}

\newcolumntype{+}{!{\vrule width 2pt}}

\newlength\savedwidth

\usepackage[aboveskip=1pt,labelfont=bf,labelsep=period,justification=raggedright,singlelinecheck=off]{caption}

\makeatletter
\renewcommand{\@biblabel}[1]{\quad#1.}
\makeatother

\date{today}

\usepackage{lastpage,fancyhdr,graphicx}
\usepackage{epstopdf}

\begin{document}
\vspace*{0.2in}

\begin{flushleft}
{\Large
\textbf\newline{Combination interventions for Hepatitis C and Cirrhosis reduction among people who inject drugs: An agent-based, networked population simulation experiment} 
}
\newline
\\
Bilal Khan\textsuperscript{1\dag},
Ian Duncan\textsuperscript{1\dag},
Mohamad Saad\textsuperscript{1\Yinyang},
Daniel Schaefer\textsuperscript{1\Yinyang},
Ashly Jordan\textsuperscript{2,3\ddag},
Daniel Smith\textsuperscript{2\ddag},
Alan Neaigus\textsuperscript{4\ddag},
Don Des Jarlais\textsuperscript{5\ddag},
Holly Hagan\textsuperscript{2,3\ddag},
Kirk Dombrowski\textsuperscript{1\dag*}
\\
\bigskip
\textbf{1} Department of Sociology, University of Nebraska, Lincoln NE
\\
\textbf{2} Rory Meyers College of Nursing, New York University, New York, NY
\\
\textbf{3} Center for Drug Use and HIV Research, New York University, New York, NY\\
\textbf{4} Department of Epidemiology, Mailman School of Public Health, Columbia University, New York, NY\\
\textbf{5} Icahn School of Medicine at Mount Sinai, New York, NY
\\
\dag Khan, Duncan and Dombrowski were the primary authors for the paper.
\ddag Jordan, Smith, Des Jarlais, Neaigus and Hagan were secondary authors and provided the majority of the data collection and analysis used to parameterize the simulation models.
\Yinyang Saad and Schaefer performed the simulation programming and contributed to the writing of the methodology.
\\
\bigskip

%
%


* corresponding author: kdombrowski2@unl.edu \\
 711 Oldfather Hall, Lincoln NE 68588, USA

\end{flushleft}
\section*{Abstract}
Hepatitis C virus (HCV) infection is endemic in people who inject drugs (PWID), with prevalence estimates above 60\% for PWID in the United States. Previous modeling studies suggest that direct acting antiviral (DAA) treatment can lower overall prevalence in this population, but treatment is often delayed until the onset of advanced liver disease (fibrosis stage 3 or later) due to cost. Lower cost interventions featuring syringe access (SA) and medically assisted treatment (MAT) for addiction are known to be less costly, but have shown mixed results in lowering HCV rates below current levels. Little is known about the potential synergistic effects of combining DAA and MAT treatment, and large-scale tests of combined interventions are rare. While simulation experiments can reveal likely long-term effects, most prior simulations have been performed on closed populations of model agents---a scenario quite different from the open, mobile populations known to most health agencies. This paper uses data from the Centers for Disease Control’s National HIV Behavioral Surveillance project, IDU round 3, collected in New York City in 2012 by the New York City Department of Health and Mental Hygeine to parameterize simulations of open populations. To test the effect of combining DAA treatment with SA/MAT participation, multiple, scaled implementations of the two intervention strategies were simulated. Our results show that, in an open population, SA/MAT by itself has only small effects on HCV prevalence, while DAA treatment by itself can significantly lower both HCV and HCV-related advanced liver disease prevalence. More importantly, the simulation experiments suggest that cost effective synergistic combinations of the two strategies can dramatically reduce HCV incidence. We conclude that adopting SA/MAT implementations alongside DAA interventions can play a critical role in reducing the long-term consequences of ongoing infection.\\
\textbf{Keywords}: hepatitis C, cirrhosis, direct acting antivirals, medically assisted opioid treatment, syringe access, simulation, intervention, network models, PWID


\section*{Introduction}
In the United States, hepatitis C virus (HCV) infection is endemic in people who inject drugs (PWID), with approximately 60\% having chronic infection, and incidence of infection among new injectors varying between 15-35 per 100 person-years of observation \cite{nelson2011epidemiology}. Beginning in 2007, HCV-related deaths in the US exceeded deaths related to HIV, and currently surpass deaths from all other notifiable infections \cite{ly2012increasing}. In high-income countries, HCV is the underlying causal factor for more than half of cases of hepatocellular carcinoma (HCC), and HCC incidence will continue to climb in the coming years due to advancing infections in the PWID (and formerly PWID) population.  There are highly effective and tolerable treatments for chronic HCV infection, but the costs of these direct-acting antiviral medications (DAAs) are high and currently fewer than 5-10\%—the estimate is imprecise---of PWID are have been treated \cite{martin2015hcv}.

Although mathematical modeling has shown that HCV treatment using the DAAs can be cost-effective \cite{iversen2017estimating,leidner2015cost}, currently 23 U.S. states require that patients have advanced HCV diagnosis (Metavir fibrosis stage F3 or cirrhosis) in order to approve publicly funded medical treatment \cite{chhatwal2016hepatitis}. In contrast, “treatment as prevention” has been proposed as a strategy to reduce prevalence among PWID---and  in turn, lower transmission \cite{hellard2015hepatitis}, though questions remain about the actual efficacy of this approach \cite{hickman2015hcv}. A systematic review and meta-analysis showed that PWID who receive medically assisted substance use treatment (SA/MAT) and participate in high-coverage syringe access programs may reduce their risk of HCV infection by 70\% \cite{platt2016effectiveness}. This raises the possibility that such strategies could be employed alongside DAA treatment to lower downstream cases of HCV-related severe liver diseases.  Previous modeling studies have reported that scaling up MAT, high-coverage syringe access programs, and HCV treatment over time can reduce new infections and disease burden, and advance toward HCV elimination \cite{martin2017modeling}. This paper extends that work by exploring the impact of treating chronically infected PWID at all HCV stages, and describing the anticipated benefits of combining DAA treatment with intensive SA/MAT.

Large-scale studies of combined interventions are rare \cite{des2010hiv}, in large part due to cost and difficulties associated with controlling recruitment and cohort retention over long time scales (as required given the slow impacts of hepatitis disease progression on health). As a result, the long-term efficacy of combination intervention strategies is largely unknown. Given the recent rapid rise in opiate abuse, data-driven intervention design is critical to altering the trajectory of HCV infection among PWID. Computer simulation of intervention effectiveness offers the possibility of predicting the long-term dynamics of HCV infection among PWID, and the opportunity to “test” \textit{in silico}, the effect of different combinations of intervention strategies \cite{martin2015hcv}. Where lower-cost SA/MAT interventions can be demonstrated to enhance the impact of DAA treatments, public health officials may be availed of new, more effective strategies to lower HCV prevalence among PWID.

Over the last two decades, researchers have used modeling and simulation to study HIV infection, uncovering important social and behavioral factors related to its prevalence among PWID \cite{morris2000microsimulation,peterson1990monte,hewitt1996spread}. This prior work has incorporated stochastic population models \cite{eaton2014health,gray2007impact,liang2016stochastic,nakagawa2016method,sloot2008stochastic} for sexual \cite{beck2015data,mei2010complex,merli2015sexual,reniers2015sexual} and injection drug co-use networks \cite{dombrowski2017interaction,khan2013network,marshall2014prevention}, and combinations of both \cite{escudero2017risk,monteiro2016understanding}. More recently, analogous approaches have been brought to bear in the context of HCV \cite{rolls2012modelling,rolls2013hepatitis}, employing simulations based on both agent-based \cite{gutfraind2015agent,wong2016parallel} and networked population perspectives \cite{hellard2015hepatitis,macgregor2017behavioural}. This ongoing effort emphasizes computational modeling of the projected impact of interventions, considered both singly \cite{hellard2015hepatitis,scott2015role} and in combination \cite{hickman2015hcv,martin2013combination}. In this paper, we adopt a network-structured, agent-based approach that combines actors with diverse behavioral profiles within the social environment of a dynamic, networked population. Our approach differs from both cohort-based Markov simulations \cite{sacks2015many} commonly used in cost-effectiveness studies of HCV treatment \cite{liu2014combining,combellick2015hepatitis} and the closed population network models found in stochastic actor simulations \cite{rolls2012modelling}. The simulations described here model open populations, where agents leave and join the risk population over time, and thus implement more realistic models of the dynamic and mobile PWID communities in which municipalities’ health departments seek to intervene. In addition, we include HIV infection in our models, as serosorting based on HIV status (and HCV status) is known to bias risk partner selection \cite{smith2013share}, and can be expected to influence the in-network epidemiological dynamics of HCV. Simulation-based research involving open, networked populations is rare, and thus the results presented here represent a significant advance over current HCV and HIV cohort modeling frameworks.

Our simulation platform has been described in detail previously \cite{khan2014stochastic}, and has been used to establish the importance of self-organizing behavioral factors in explaining the non-spreading of HIV among PWID in New York City during the early stages of HIV epidemic \cite{khan2013network,dombrowski2017interaction}. In the current analysis, simulations track  long-term HCV infection rates under a range of conditions: 1) baseline---where no new prevention or treatment activities are involved; 2) scaled implementation of MAT and intensified syringe access to reduce pair-wise risk of infection); 3) implementation of DAA treatment to chronically infected agents; and 4) combinations of the aforementioned prevention and treatment strategies at a range of implementation scales. Throughout, intervention results are tracked over 15 years from an initial starting point that is based on parameters drawn from PWID surveillance research in New York City \cite{neaigus2017trends,jordan2015incidence}.

\section*{Methods}
\subsection*{Data}
Data for simulations is drawn from the Centers for Disease Control’s National HIV Behavioral Surveillance project \cite{wejnert2017achieving}, IDU round 3, collected in New York City in 2012 by the New York City Department of Health and Mental Hygiene \cite{ivy2015hiv,neaigus2017trends}. Table \ref{univariates}, located in the Supporting Information (below), provides the univariate distributions drawn from participants who were verified to have injected any drug in the past year at the time of their interview. These data were used to create each individual agent’s age, gender, average network degree, race, HIV status, HCV status, and longevity of participation within the simulation population.

In the simulation, risk relationships were modeled as network edges.  Tables showing the bivariate distributions that quantify the homophily biases on population mixing (by age, average degree--binned, gender, HCV status, HIV status, and race) are listed in the Supporting Information, below (see Table \ref{bivariates}). To remain true to the original recruitment data, these distributions often reflect asymmetrical mixing tendencies common to PWID---meaning that the tendency of individuals in group A to seek partners from group B may not be the same as the tendencies of B’s to seek partners from A. The parameters in these tables were drawn from both the interview data and the respondent driven sampling recruitment data, analyzed and weighted via RDSAT \cite{spiller2012rds}. Tendencies toward serosorting behaviors for both HIV and HCV infection are drawn from published sources \cite{burt2009serosorting,smith2013share,duncan2017hepatitis}. Prior research has shown that serosorting factors condition the likelihood of pairwise risk relationships \cite{dombrowski2013reexamination}, and the use of serosorting data has yielded conclusions which are consistent with what is known about the long-term behaviors of the PWID population \cite{khan2013network} in New York City. Distributions quantifying the duration of risk relationships were drawn from consultation with ethnographers working in the area and prior research with PWID in New York City \cite{khan2013network} .

Risk act frequencies were drawn from the interview portion of the NHBS survey (approximately one per every two to three weeks; from NHBS variables ``injavgr; sharndler; sharcookr; sharcottr; sharwatrr; sharworkr; samessyrr'').  Per-risk-act infection probabilities were treated as a tuning parameter. The final setting (0.009\% chance of infection per risk act across discordant pairs in an active risk relationship) produced prevalence levels (60-70\%) that match current HCV levels in the NHBS New York PWID population and among US PWID more generally \cite{neaigus2017trends,nelson2011epidemiology}---see Boelen et al \cite{boelen2014per} for a discussion of defining transmission probability via equipment sharing. HCV pathogen properties of spontaneous clearance \cite{smith2016spontaneous} and stage transition probabilities (e.g. Metavir fibrosis stage advancement rates and the development of decompensated cirrhosis, and hepatocellular carcinoma) were drawn from published meta-analyses of HCV infection dynamics among PWID \cite{smith2015hepatitis}.

\subsection*{Simulation}
In what follows, each distinct intervention strategy is referred to as a “scenario”.  For each scenario considered, 5 random artificial injection networked populations of 10,000 injectors were created.  For each of these networks, 3 simulations were undertaken, projecting the population out 20 years into the future. As described above, the initial demographics, relational structure, and behavioral mechanics of these synthetic populations was based on statistics derived from NYC NHBS data. The initial prevalence of chronic HCV infection for the population was set at ~50\%, meaning that, at the beginning of the simulation, 50\% of the population is randomly assigned an HCV infection state according to a distribution drawn from published sources (these data were not available in the NHBS data obtained for this study). After 5 years of simulation “burn in”, HCV prevalence stabilizes at 60-65\%, and HIV prevalence stabilizes at 12-15\% \cite{neaigus2017trends}. Such conditions are meant to approximate current conditions in New York City, and generally reflect PWID HCV prevalence rates in urban areas through the US at the current time. The overall population levels for the simulation (i.e. 10,000 agents) is meant to reflect a single, open region within a larger urban zone. Such geographical concentrations are well known in most urban areas and represent permeable spatial and social foci where individual PWID are associated for varying lengths of time \cite{curtis1995street}. The burn in period is meant to produce a simulation environment that is free from initial starting conditions. Generally, as will be seen below, HCV prevalence during the burn in period rises as HCV is transmitted across the risk edges to uninfected agents.

Over the course of each scenario, agents commit risk acts across their existing risk relationships, and form new partnerships whenever their existing connections dissolve. Agents also leave the simulation population when their participation longevity elapses. Whenever an agent leaves the simulation, his/her network connections are dissolved and his/her HCV infection status is removed from future measurements. To replace departing agents, new agents are created and form risk partnerships (with biases dictates by the bivariate distribution, conditioned on their own demographics and behavioral attributes). Arriving agents continue to have a 50\% chance of being HCV infected at inception (as per initial conditions---simulating a situation where a mixture of new injectors and those who have been injecting for some time enter the population from the surrounding area). At each step of the simulation, population-wide prevalence is measured by considering the status of only those agents who are still in the simulation. This modeling paradigm reflects real-world conditions where community boundary conditions are fluid and populations are mobile: PWID change locations, enter addiction treatment, become incarcerated, and so on, all of which removes them from the simulation, while other, previously unknown PWID take their place in an evolving risk network milieu (54). Individual infections of both HIV and HCV also age as the simulation progresses. Metavir stages are incremented for those infected with HCV, and stochastic opportunities for disease progression are made each year (with probabilities set to match known population level outcomes). Among the possible outcomes is death from infection by HCV or HIV/HCV co-infection via liver-related disease \cite{smit2008risk}.

\subsection*{Interventions}
In the experiments described here, we test the effect of a range of intervention scenarios:
\begin{itemize}
\item A baseline scenario where no interventions take place. 
\item A DAA intervention implemented on the first day of year 6 (following the 5-year burn in period) and on the first day of each subsequent year. Here a fixed percentage of the current, chronically infected HCV population is randomly selected for participation in the intervention, and their start date is randomly assigned to occur in the next 365 days. During the intervention, participants are shielded from risk events and undergo treatment for HCV. Reflecting research findings, 90\% of those agents adhere to HCV DAA treatment for the entire duration (168 days), and 95\% of those who complete it are cured \cite{lawitz2013sofosbuvir}. At the end of treatment adherence period agents are placed back into the simulation population and continue to operate in a manner specified by their individual agent parameters.
\item Intensive syringe access and opiate substitution treatment (SA/MAT), at various levels of recruitment/participation. Here, in year 6 and in each year following, a fixed percentage of the total population in the simulation is recruited for participation regardless of HCV status. During the year that follows, each participating agent’s number of risk acts with their network neighbors is decreased by 80\% from its original pre-intervention level \cite{turner2011impact}. At the end of the intervention period, participating agents return to their pre-intervention risk act rate. Because the selection of participants is random each year, individuals may be re-recruited in subsequent years (or even two or more years in a row), depending on the luck of the draw; or they may leave the intervention at any time depending on their likelihood of completing the program (determined stochastically). The result is a distribution of participation that may last from a few months to several years. In cases where a high percentage of the population is enrolled each year, the latter becomes more likely.
\end{itemize}
Other than the changes in risk behavior (in SA/MAT participants) or change of disease status (in DAA participants), agent behaviors remained consistent with baseline.

To test the effect of combination strategies, additional scenarios were considered where both the DAA treatment and SA/MAT participation were engaged concurrently (and independently) in the same simulation population.  The rationale for this set of scenarios was to determine whether and to what extent each of these approaches---direct-acting antivirals and opiate substitution/syringe access---might, in combination, further reduce levels of HCV in the population compared against approaches relying on a single strategy and a baseline scenario matching current day conditions.

\subsection*{Validation}
The results of the simulation are shown as averages and standard deviations across the 15 runs for each simulation scenario. Whereas simulations begin from a random network and all simulation events are determined by stochastic processes, results that appear consistently across a number of simulations are interpreted as highly likely. Robustness and sensitivity tests for the simulations are discussed in the Limitations section below.  Validation of the Baseline model can be seen in the incidence (per 100 person years) and prevalence of the main outcome variables: acute HCV infection, chronic HCV infection (Metavir fibrosis stages 0-3), cirrhosis (CC), decompensated cirrhosis (DC), and hepatocellular carcinoma (HCC) (see Table \ref{baseline}). These results can be compared to those of recent meta-analyses \cite{smith2015hepatitis} and NYC incidence and prevalence estimates \cite{jordan2015incidence,neaigus2017trends}.

\begin{table}[htb]
\begin{adjustwidth}{0.25in}{0in} 
\centering
\caption{\bf Average incidence and prevalence of HCV states in baseline simulations}
\begin{tabular}{lcccccc}
\hline
HCV State & AVG Incd. & std & Pub. Rate & Avg Prev. (\%) & std (\%) & Pub. Rate\\ \hline
Uninfected	& & & & 33.9 & 2.1 & \\ \hline
Acute &	11.841 & 0.420 & & 4.6 & 0.2 & \\ \hline
CC & 0.599 & 0.023 & 0.662 & 3.9 & 0.5 & \\ \hline
DC & 0.088 & 0.006 & 0.182 & 0.8 & 0.1 & \\ \hline	
HCC & 0.024 & 0.002 & 0.032 & 0.5 & 0.1 & \\ \hline	
All HCV States & & & & 66.1 & 1.8 & 67.1 \\ \hline
\end{tabular}
\begin{flushleft}\textit{Measures after the completion of 15 years simulation that followed a 5-year burn-in period. Results show the mean and standard deviation for each state from 15 independent simulations of networked populations of 10,000 actors. Acute incidence expressed per 100 person years of all uninfected agents. Cirrhosis (CC), Decompensated Cirrhosis (DC), and Hepatocellular (HCC) Carcinoma incidence expressed per 100 person years of all HCV infected agents. For published incidence and prevalence rates see Jordan et al \cite{jordan2015incidence} and Neaigus et al \cite{neaigus2017trends}}.
\end{flushleft}
\label{baseline}
\end{adjustwidth}
\end{table}

\section*{Results}
The final prevalence rates for the 15 years after burn in (for Baseline) and after burn in and intervention implementation (for a range of DAA and SA/MAT interventions scales) are found in Table \ref{interventions}. They show similarities to previous results from Europe \cite{martin2015hcv}. In particular, SA/MAT shows limited ability to lower overall HCV prevalence (declines of 3-12\%)---and very little detectable effect on advanced liver disease after 15 years \cite{holtzman2009influence}. This is not surprising, given that SA/MAT does not cure infection---and thus can not affect prevalence rates directly. However, as reflected in the declining prevalence of acutely infected agents, SA/MAT shows a steady decline in HCV incidence as the scale of the intervention increases. Here, as the population turns over---with formerly infected agents leaving the network and new, uninfected agents arriving---prevalence rates can drop due to lowered infection rates. Scaled DAA treatment shows the ability to lower HCV rates directly, resulting in a more than 30\% decline when 20\% of those with chronic HCV infection are treated each year. Here, in contrast to SA/MAT intervention, there is very little effect on incidence of acute infection however. The simulations also show a large drop in advanced liver disease associated with DAA treatment over the 15-year intervention period (reduction of more than 40\%). These results were obtained despite the unbounded nature of the population and the fact that many of those who undergo DAA treatment subsequently leave the simulation in the future and are replaced by others who did not undergo treatment.

\begin{table}[htb]
\begin{adjustwidth}{0.25in}{0in} 
\centering
\caption{\bf Population proportion by HCV disease state after 15 years of intervention: mean (std)}
\begin{tabular}{lccccc}
\hline
Intervention & Uninfected & Acutely infected & F0-3 & CC+DC+HCC & Total \\ \hline
Baseline & .384 (.005) & .049 (.002) & .527 (.005) & .041 (.002) & .616 \\ \hline
DAA @ 5 & .440 (.004) & .051 (.002) & .475 (.004) & .035 (.002) & .560 \\ \hline
DAA @ 10 & .491 (.006) & .051 (.003) & .427 (.005) & .031 (.002) & 	.509 \\ \hline
DAA @ 15 & .535 (.004) & .052 (.002) & .390 (.004) & .027 (.002) & .465 \\ \hline
DAA @ 20 & .574 (.004) & .049 (.002) & .352 (.004) & .024 (.001) & .426 \\ \hline
SA/MAT @ 20 & .400 (.008) & .045 (.003) & .514 (.007) & .041 (.002) & .600 \\ \hline
SA/MAT @ 40 & .417 (.005) & .040 (.002) & .503 (.005) & .040 (.002) & .583 \\ \hline
SA/MAT @ 60 & .438 (.005) & .034 (.002) & .489 (.006) & .039 (.002) & .562 \\ \hline
SA/MAT @ 80 & .457 (.005) & .028 (.002) & .475 (.006) & .040 (.002) & .543 \\ \hline
\end{tabular}
\begin{flushleft} \textit{Measures after the completion of 15 years simulation that followed a 5-year burn-in period. Results show the mean and standard deviation for each state from 15 independent simulations of networked populations of 10,000 actors.}
\end{flushleft}
\label{interventions}
\end{adjustwidth}
\end{table}

The long-term effects of combined interventions at various scales are considered next, including measurements of chronic prevalence (Table \ref{chronic}), prevalence of HCV-related advanced liver disease (CC+DC+HCC; see Table \ref{severe}), HCV incidence (Table \ref{hcvinc}), and incidence of cirrhosis (Table \ref{cirrinc}). The tabulated results show that the effects of SA/MAT treatment marginally enhance the effectiveness of DAA treatment in lowering chronic HCV prevalence in an open population, but have little effect on HCV-related advance liver disease, with roughly linear effects due to scale. 

\begin{table}[htb]
\begin{adjustwidth}{0.25in}{0in} 
\centering
\caption{\bf Population proportion of Chronic (F0-F3) after 20 years of combined interventions}
\begin{tabular}{lccccc}
\hline
Intervention & Baseline & DAA @ 5 & DAA @ 10 & DAA @ 15 & DAA @ 20 \\ \hline
Baseline & .527 (.005) & .475 (.004) & .427 (.005) & .390 (.004) & .352 (.004) \\ \hline
SA/MAT @ 20 & .514 (.007) & .477 (.006) & .413 (.003) & .377 (.004) & .343 (.004) \\ \hline
SA/MAT @ 40 & .503 (.005) & .453 (.007) & .403 (.004) & .366 (.005) & .332 (.004) \\ \hline
SA/MAT @ 60 & .489 (.006) & .436 (.006 & .391 (.005) & .355 (.004) & .323 (.005) \\ \hline
SA/MAT @ 80 & .475 (.006) & .423 (.006) & .378 (.005) & .345 (.004) & .313 (.004) \\ \hline
\end{tabular}
\begin{flushleft} \textit{Measures after the completion of 15 years simulation that followed a 5-year burn-in period. Results show the mean and standard deviation for each state from 15 independent simulations of networked populations of 10,000 actors. Acute infections are not included in this figure to avoid counting those who spontaneous clear the infection during the simulation year.}
\end{flushleft}
\label{chronic}
\end{adjustwidth}
\end{table}

\begin{table}[htb]
\begin{adjustwidth}{0.25in}{0in} 
\centering
\caption{\bf Population proportion of HCV-related advanced liver disease (CC, DC, HCC) after 15 years of combined interventions}
\begin{tabular}{lccccc}
\hline
Intervention & Baseline & DAA @ 5 & DAA @ 10 & DAA @ 15 & DAA @ 20 \\ \hline
Baseline & .041 (.002) & .035 (.002) & .031 (.002) & .027 (.002) & .024 (.001) \\ \hline
SA/MAT @ 20 & .041 (.002) & .034 (.002) & .031 (.001) & .027 (.002) & .023 (.001) \\ \hline
SA/MAT @ 40 & .040 (.002) & .035 (.003) & .030 (.002) & .026 (.002) & .023 (.002) \\ \hline
SA/MAT @ 60 & .039 (.002) & .033 (.002) & .029 (.001) & .025 (.001) & .023 (.001) \\ \hline
SA/MAT @ 80 & .040 (.002) & .034 (.002) & .030 (.002) & .025 (.002) & .023 (.002) \\ \hline
\end{tabular}
\begin{flushleft} \textit{Measures after the completion of 15 years simulation that followed a 5-year burn-in period. Results show the mean and standard deviation for each state from 15 independent simulations of networked populations of 10,000 actors.}
\end{flushleft}
\label{severe}
\end{adjustwidth}
\end{table}

\begin{table}[htb]
\begin{adjustwidth}{0.25in}{0in} 
\centering
\caption{\bf  Incidence rate of HCV among uninfected agents in 100 person years)}
\begin{tabular}{lccccc}
\hline
Intervention & Baseline & DAA @ 5 & DAA @ 10 & DAA @ 15 & DAA @ 20 \\ \hline
Baseline & 11.00 (0.25) & 10.40 (0.35) & 9.37 (0.39) & 8.84 (0.28) & 7.89 (0.33) \\ \hline
SA/MAT @ 20 & 9.43 (0.68) & 8.72 (0.30) & 8.07 (0.28) & 7.35 (0.16) & 6.74 (0.32) \\ \hline
SA/MAT @ 40 & 7.74 (0.37) & 7.44 (0.67) & 6.33 (0.33) & 5.90 (0.21) & 5.29 (0.26) \\ \hline
SA/MAT @ 60 & 6.02 (0.38) & 5.58 (0.41) & 5.02 (0.32) & 4.46 (0.28) & 4.16 (0.28) \\ \hline
SA/MAT @ 80 & 4.38 (0.29) & 4.16 (0.30) & 3.57 (0.28) & 3.23 (0.20) & 2.93 (0.15) \\ \hline
\end{tabular}
\begin{flushleft} \textit{Measures after the completion of 15 years simulation that followed a 5-year burn-in period. Results show the mean and standard deviation for each state from 15 independent simulations of networked populations of 10,000 actors.}
\end{flushleft}
\label{hcvinc}
\end{adjustwidth}
\end{table}

\begin{table}[htb]
\begin{adjustwidth}{0.25in}{0in} 
\centering
\caption{\bf Incidence rate of Cirrhosis among all infected HCV agents (in 100 person years)}
\begin{tabular}{lccccc}
\hline
Intervention & Baseline & DAA @ 5 & DAA @ 10 & DAA @ 15 & DAA @ 20 \\ \hline
Baseline & .554 (.040) & .531 (.076) & .500 (.060) & .450 (.067) & .405 (.065) \\ \hline
SA/MAT @ 20 & .570 (.099) & .521 (.081) & .474 (.044) & .435 (.070) & .390 (.062) \\ \hline
SA/MAT @ 40 & .521 (.053) & .512 (.070) & .473 (.059) & .406 (.076) & .370 (.068) \\ \hline
SA/MAT @ 60 & .552 (.047) & .473 (.072 & .432 (.057) & .399 (.079) & .385 (.050) \\ \hline
SA/MAT @ 80 & .515 (.063) & .479 (.066) & .435 (.065) & .381 (.054) & .339 (.057) \\ \hline
\end{tabular}
\begin{flushleft} \textit{Measures after the completion of 15 years simulation that followed a 5-year burn-in period. Results show the mean and standard deviation for each state from 15 independent simulations of networked populations of 10,000 actors.}
\end{flushleft}
\label{cirrinc}
\end{adjustwidth}
\end{table}

A different pattern emerges when we turn our attention to incidence rates.  Here we see that SA/MAT interventions have significant impacts. As seen in Table \ref{hcvinc}, high levels of SA/MAT recruitment (at 80\% treatment level) have the effect of lowering overall HCV incidence by more than 60\%, while scaled DAA treatment at even the highest levels produces a less significant effect (a decrease of 28\%). Even at low intervention levels (20\%), SA/MAT treatment lowers HCV incidence by 21-39\% when combined with DAA treatment, with the impact increasing as DAA intervention levels rise. SA/MAT interventions have less of an impact on cirrhosis incidence associated with HCV infection, but still serve to lower cirrhosis incidence by 6-30\% at even low levels. As seen in Table \ref{cirrinc}, these impacts continue to grow as SA/MAT increases in scale of coverage.

\section*{Discussion}
This simulation experiment represents an attempt to model open populations of PWID over long time periods, and to test the effectiveness of single and combined interventions under such conditions. The scenario we envision is quite common to municipalities and public health departments facing rising HCV rates associated with increases in injection drug use: what are the best strategies for reducing HCV incidence and prevalence and subsequent severe liver related disorders? Given the high mobility of PWID and the long time periods needed for intervention impacts to manifest, intervention planners need to know what can be expected of new disease containment strategies based on the administration of direct acting antivirals, the deployment of medically-assisted addiction treatment and high-coverage syringe access programs, and combinations thereof.

The open nature of the simulations added considerable dynamism to a model that already included regular network ``churn''---the effect of agents leaving some risk relationships and entering into relationships with new alters. Despite the fact that the simulation population was held steady at a level of 10,000 agents, in each 20-year simulation: more than 28,000 agents participated in the simulation; more than 330 thousand risk relationships were made and unmade; more than 131,000 intervention events took place; and approximately 26,000 total new infections (HIV and HCV) occurred. The effects of overall HCV incidence and prevalence we report here were thus consistent even amid considerable network change and population turnover.

The results of the simulation verify previous research findings that SA/MAT interventions, acting alone, do little to reduce overall prevalence of HCV infection or the related advanced liver diseases \cite{tsui2014association}. However, our result show that SA/MAT does have a significant effect on HCV incidence---an effect that, when coupled with even low levels of DAA treatment, can magnify the much smaller declines in HCV incidence under DAA treatment alone. The long term effects of this are shrouded by the very long time scales associated with advancing HCV, and by the open boundaries on the simulation community---whereby the effects of treatment on the network are lost when those treated leave the simulation. Low rates of DAA (5\% annually) can be coupled with moderate SA/MAT (40\%) to produce a 32\% drop in HCV incidence and a 8\% drop in cirrhosis incidence over 15 years. In contrast, the same DAA treatment that takes place without accompanying SA/MAT shows smaller declines in HCV incidence (4\%), ensuring that treatment needs remain high in the future. High levels of DAA (20\%) and SA/MAT (80\%) in combination show an ability to lower both HCV and cirrhosis incidence by more than 73\% and 39\%, respectively, when compared to baseline levels. Importantly, these effects can be anticipated despite high population turnover, and take place amid a steep drop in HCV prevalence.

\begin{figure}[tb]
\centering
\includegraphics[width=5.0in]{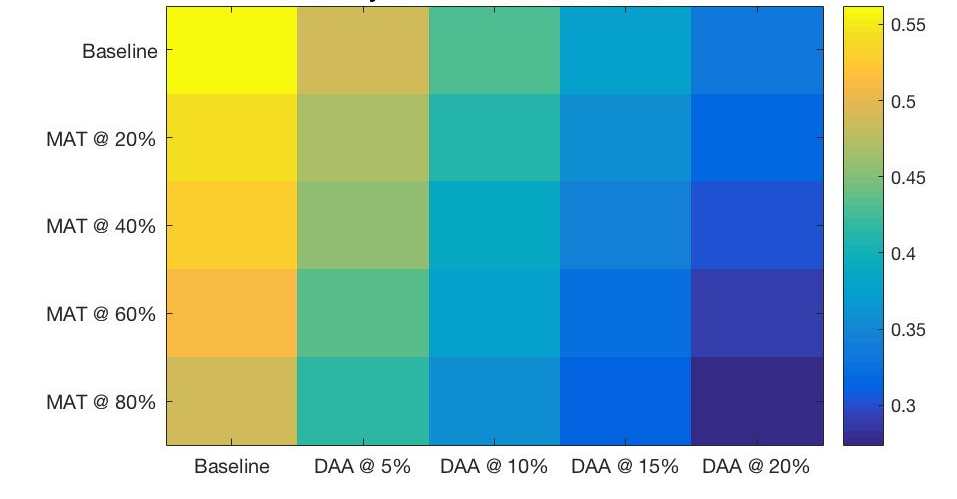}
\caption{\textit{Chronic HCV (F0-3) prevalence at a range of combined intervention scales after 15 years of simulated intervention. Thresholds can be seen at the DAA 20\% scale for all levels of SA/MAT.}}
\label{figure1}
\end{figure}

\begin{figure}[t]
\centering
\includegraphics[width=5.0in]{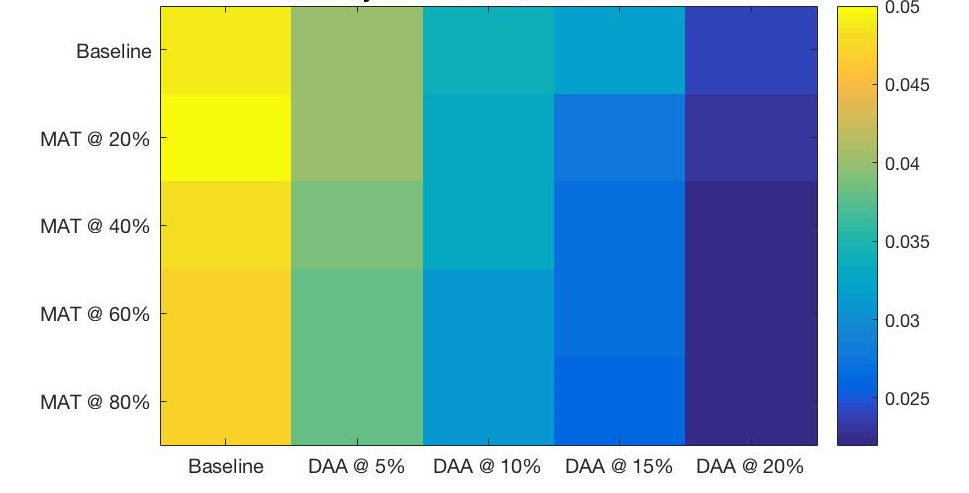}
\caption{\textit{Advanced HCV-related liver disease prevalence at a range of combined intervention scales after 15 years of simulated intervention. Strong thresholds can be seen at the DAA 20\% scale for all levels of SA/MAT.}}
\label{figure2}
\end{figure}

\begin{figure}[htb]
\centering
\includegraphics[width=5.0in]{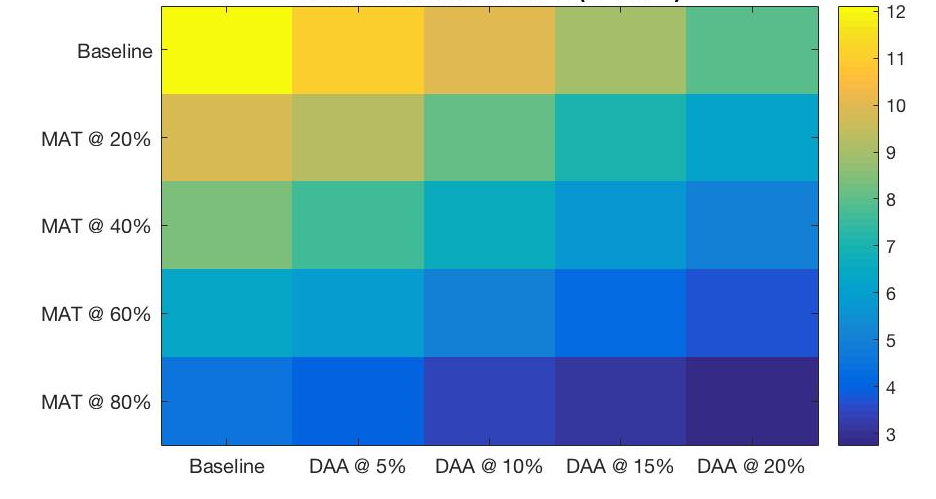}
\caption{\textit{HCV incidence (in 100 person years) at a range of combined intervention scales after 15 years of simulated intervention. Strong thresholds can be seen at the SA/MAT 80\% scale for all levels of DAA treatment, and for SA/MAT 60\% at DAA treatments of 10\% or higher.}}
\label{figure3}
\end{figure}

\begin{figure}[t]
\centering
\includegraphics[width=5.0in]{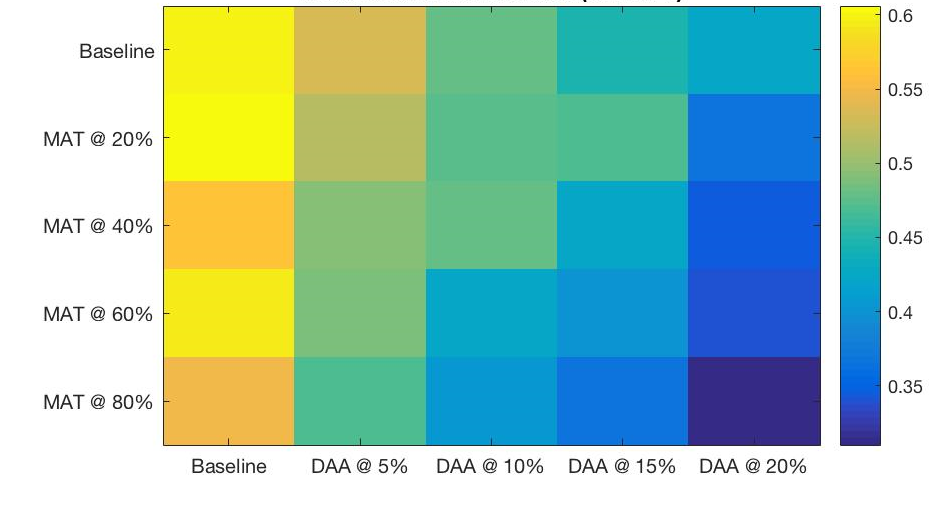}
\caption{\textit{Cirrhosis incidence at a range of combined intervention scales after 15 years of simulated intervention. A notable threshold can be seen at the DAA 20\% and SA/MAT 20\% scale. There the inclusion of even low levels of SA/MAT can combine with higher levels of DAA to produce considerable effects on cirrhosis incidence in later years.}}
\label{figure4}
\end{figure}

The heatmaps shown in Figures \ref{figure1} -- \ref{figure4} demonstrate both continuous and thresholding behaviors in the combined interventions. Chronic HCV prevalence (Figure \ref{figure1}) and the prevalence of advanced liver diseases (Figure \ref{figure2}) show continuous declines as DAA levels increase, with relatively fewer (but still continuous) declines provided by increased levels of SA/MAT. The major threshold in these results arrives at the 20\% DAA treatment level, particularly for advanced liver diseases. In contrast, incidence levels for HCV (Figure \ref{figure3}) show thresholds at SA/MAT 60 for moderate (10\%) DAA treatments. At SA/MAT 80, a strong threshold exists for all levels of DAA treatment. Similarly, one notes a strong threshold in cirrhosis incidence (Figure \ref{figure4}) at the DAA 20-SA/MAT 20 level. These thresholds imply an opportunity for discontinuous improvements in prevention effects at these levels. In a DAA 20 intervention, for example, considerable declines in downstream cirrhosis incidence can be achieved by adding even low level SA/MAT interventions (20\%) now. Overall HCV incidence can be lowered dramatically by including high levels of SA/MAT (80\%) at all scales of DAA treatment.

\subsection*{Limitations}
The most significant limitation on these conclusions is the obvious fact that these results were obtained via simulation, and thus ignore a range of possible influences external to the simulation environment---such as changes in drug markets and drug availability, the impact of possible historical events (such as the long term implications of the unfolding opioid epidemic in the US), changes in PWID routes of administration \cite{neaigus2006transitions}, or local efforts aimed at changing risk behaviors. Though our framework attempts to engage more realistic intervention conditions through the simulation of an open, networked population, possible errors in these data and their basis in New York City also limit their possible extension to other locations.

To test the robustness of our results against possible sensitivity issues unanticipated in the modeling framework, we conducted sensitivity tests that examined significant changes in input parameters closely connected to HCV spreading in the population. Using “one-factor-at-a-time” strategies \cite{czitrom1999one}, we altered risk rates, HCV transmission probabilities, network churn rates, and initial prevalence levels of both HIV and HCV (see Table \ref{sensitivity}). Perturbations of 15\% (up and down) from initial levels showed minimal impact on HCV prevalence after 15 years in both baseline and intervention scenarios for all factors except initial HCV prevalence. The sensitivity of final HCV prevalence to initial prevalence levels is not surprising given the open population model and the fact that new agents entering the simulation take their HCV status probabilistically from the initial settings. Still, a 15\% perturbation in initial HCV rates resulted in a final HCV prevalence change of only 7-9\% in the baseline scenarios, and 11-12\% in the intervention scenarios. In effect, the overall impact of the change was blunted by the simulation process. From these tests, we surmise that our results are robust against small to medium-sized errors in initial settings, and that the results are unlikely to reflect hidden “butterfly” effects.

\begin{table}[b]
\begin{adjustwidth}{0.25in}{0in} 
\centering
\caption{\bf Sensitivity test showing the effects of parameter changes of 15\% on HCV prevalence under both baseline and intervention conditions}
\begin{tabular}{lcccc}
\hline
& Baseline & (orig. 0.616) & DAA 10/MAT 40 & (orig. 0.473) \\ \hline
& +15\%	& -15\% & +15\% & -15\% \\ \hline
Churn rate & 0.613 (0.005) & 0.618 (0.005) & 0.467 (0.004) & 0.476 (0.010) \\ \hline
Risk interval mean & 0.590 (0.007) & 0.630 (0.008) & 0.445 (0.005) & 0.4814 (0.002) \\ \hline
Transmission probability & 0.638 (0.006) & 0.596 (0.008) & 0.483 (0.007) & 0.467 (0.011) \\ \hline
Initial HIV prev. & 0.617 (0.005) & 0.615 (0.006) & 0.475 (0.008) & 0.478 (0.003) \\ \hline
Initial HCV prev. & 0.671 (0.002) & 0.570 (0.006) & 0.532 (0.006) & 0.420 (0.004) \\ \hline
\end{tabular}
\begin{flushleft} \textit{Measures after the completion of 15 years simulation that followed a 5-year burn-in period. Results show the mean and standard deviation for each state from 5 independent simulations of networked populations of 10,000 actors.}
\end{flushleft}
\label{sensitivity}
\end{adjustwidth}
\end{table}

\section*{Conclusion}

These results suggest significant and immediate steps for health officials and harm reduction programs. Combined interventions that match lower-cost syringe access and medicine assisted addiction treatments with direct acting antiviral treatments can have a large effect on HCV incidence, cirrhosis incidence, and overall HCV prevalence, even among highly mobile PWID populations. Simulations suggest that lower cost syringe exchange and medically assisted opioid treatment can be combined with less available direct acting antiviral treatments to radically lower HCV incidence among people who inject drugs, even when that population is highly mobile and where treatment is at times short term and only temporarily successful. These outcomes are contingent on opening up participation in direct acting antiviral treatment to drug users prior to their entry into later stage fibrosis, and on the ability of health agencies to sustain intervention efforts for more than a decade. 

\section*{Acknowledgments}
The HCV Synthesis Project is supported by a grant from the National Institutes of Health (RO1DA034637). Support was also received from the New York University Center for Drug Use and HIV Research, NIH P30 Center (P30DA011041). The modeling research presented here is supported by NIH/NIDA (R01DA037117). The original simulation platform was developed under RC1DA028476. Data were obtained from the New York City Department of Health and Mental Hygiene, HIV Epidemiology and Field Services through a cooperative agreement. Special thanks to Sarah Braunstein, Director of HIV Epidemiology at NYC DOHMH for this collaboration. Funding for the data collection was provided via a cooperative agreement between the Centers for Disease Control and Prevention, National HIV Behavioral Surveillance program and the New York City Department of Health and Mental Hygiene, Grant U62/CCU223595-03-1. For the 2012 NHBS IDU Round 3 data collection, Alan Neaigus, MCRP, PhD, served as Principal Investigator and Kathleen H. Reilly, PhD, MPH, served as Project Director. Their work, and those of the NHBS NYC IDU Round 3 field team---Holly Hagan, Travis Wendel, David Marshall III, Roberto Abadie, and Carmen Anna Davila---is acknowledged and appreciated. 

\newpage
\bibliography{hcvdaa}

\section*{Supporting information}
Simulation design and stochastic processes are described elsewhere \cite{khan2014stochastic} . The parameters for the current simulations were drawn from NHBS IDU Round 3 study in New York City. Data collected in 2012 were used to create both agent characteristics and a range of network mixing parameters. Univariate distributions for creating simulation agents can be found in Table \ref{univariates}. All variables are treated as categorical to enable discrete mixing probabilities. Where ranges are present within a category (i.e. ``age'' or ``average degree''), agents are assigned a random value within the range. 

\begin{table}[h]
\centering
\caption{\bf Per agent univariate parameters}
\begin{tabular}{llcc}
\hline
Variable & Source & Value & Percentage \\ \hline
Age & NHBS--``age'' & 18-24 & 1.0 \\ \hline
& & 25-34 & 16.7 \\ \hline
& & 35-44 & 28.1 \\ \hline
& & 45-54 & 38.4 \\ \hline
& & 55-65 & 13.9 \\ \hline \\
Average Degree & NHBS (see below) & 0-3 & 62.1 \\ \hline
& & 4-8 & 18.4 \\ \hline
& & 9-19 & 19.5 \\ \hline \\
Connection Duration & Khan et al \cite{khan2014stochastic} & Short & 40 \\ \hline
& & Long & 60 \\ \hline \\
Gender & NHBS--``gender'' & Male & 74.9 \\ \hline
& &	Female & 24.5 \\ \hline 
& & Transgender & 0.1 \\ \hline \\
HCV initial & (Ashly…source?) & Negative & 50 \\ \hline
& & Acute & 10.0 \\ \hline
& & Chronic (F0) & 30.0 \\ \hline
& & Fibrosis Stage 1 & 4.5 \\ \hline
& & Fibrosis Stage 2 & 2.0 \\ \hline
& & Fibrosis Stage 3 & 1.5 \\ \hline
& & Cirrhosis & 1.0 \\ \hline
& & Decompensated & 0.5 \\ \hline
& & HCC & 0.5 \\ \hline \\
HIV & Neaigus et al \cite{neaigus2017trends} & Uninfected & 88.0 \\ \hline
& & Acute & 0.2 \\ \hline
& & Chronic & 11.8 \\ \hline \\
Longevity & NHBS--``age-ageinj'' & 7-30 days & 1 \\ \hline
& & 365-7301 days & 99 \\ \hline \\
Race/Ethnicity & NHBS--``newrace'' & African American & 66.7 \\ \hline
& & Hispanic & 13.5 \\ \hline
& & White & 18.8 \\ \hline
& & Other & 1.0 \\ \hline
\end{tabular}
\begin{flushleft} \textit{Baseline agent parameter distributions. The degree distribution was the sum of the NHBS variables ``num-na; num-ccw; num-dda''. This count provides the number of risk partners---i.e. the expected number of network alters with whom an agent shares needles or other injection related equipment. This number should not be confused with the common measure of RDS network degree often used in data collection with PWID (which is the number of known alters who also inject drugs). The latter is normally much higher than the actual number of risk partners.}
\end{flushleft}
\label{univariates}
\end{table}

Mixing patterns between agents were determined by aggregating the agent x agent bivariate probabilities across all categories to create a relatively likelihood of connection from agent to agent (see Table \ref{bivariates}). Stochastic throws against relative likelihoods were used to determine eventual connections. Bivariate distributions were calculated using the recruitment sampling data from the New York City NHBS IDU Round 3 data, weighted and adjusted using RDSAT analysis.

\begin{table}[htb]
\centering
\caption{\bf Per agent bivariate parameters}
\begin{tabular}{lccccc}
\hline
\textbf{Age (binned)} & & & & & \\ \hline
& 18-24 & 25-34 & 35-44 & 45-54 & 55-65 \\ \hline
18-24 & 0.0 & 0.525 & 0.273 & 0.202 & 0.0 \\ \hline
25-34 & 0.083 & 0.248 & 0.302 & 0.255 & 0.075 \\ \hline
35-44 & .027 & 0.184 & 0.331 & 0.368 & 0.091 \\ \hline
45-54 & 0.014 & 0.110 & 0.173 & 0.443 & 0.174 \\ \hline
55-65 & 0.0 & 0.087 & 0.173 & 0.471 & 0.269 \\ \hline \\
\textbf{Ideal Degree} & & & & & \\ \hline
& 0-3 & 4-8 & 9-19 & & \\ \hline
0-3 & 0.518 & 0.364 & 0.118 & & \\ \hline
4-8	& 0.391 & 0.417 & 0.193 & & \\ \hline
9-19 & 0.279 & 0.425 & 0.296 & & \\ \hline \\
\textbf{Gender} & & & & & \\ \hline
& Male & Female & Trans. & & \\ \hline
Male & 0.784 & 0.211 & 0.005 & & \\ \hline	
Female & 0.644 & 0.345 & 0.011 & & \\ \hline 
Transgender & 0.379 & 0.288 & 0.333 & & \\ \hline \\
\textbf{HCV Status} & & & & & \\ \hline
& Positive & Negative & & & \\ \hline	
Positive & 0.814 & 0.186 & & & \\ \hline			
Negative & 0.286 & 0.714 & & & \\ \hline \\	
\textbf{HIV Status} & & & & & \\ \hline
& Positive & Negative & & & \\ \hline				
Positive & 0.816 & 0.184 & & & \\ \hline				
Negative & 0.144 & 0.856 & & & \\ \hline \\			
\textbf{Race/Ethnicity} & & & & & \\ \hline	
& Afr. Amer. & Hispanic & White & Other & \\ \hline	
African American & 0.762 & 0.100 & 0.127 & 0.010 & \\ \hline
Hispanic & 0.507 & 0.381 & 0.112 & 0.0 & \\ \hline
White & 0.277 & 0.048 & 0.651 & 0.0 & \\ \hline		
Other & 0.472 & 0.0 & 0.528 & 0.0 & \\ \hline
\end{tabular}
\begin{flushleft} \textit{The table should be read as the proportional probability for each variable of forming an outgoing tie from the row to the column.}
\end{flushleft}
\label{bivariates}
\end{table}


\end{document}